# Calibration of photomultipliers gain used in the J-PET detector

T. Bednarski[1], E. Czerwiński[1], P. Moskal[1], P. Białas[1], K. Giergiel[1], Ł. Kapłon[1,2], A. Kochanowski[2], G. Korcyl[1], J. Kowal[1], P. Kowalski[3], T. Kozik[1], W. Krzemień[1], M. Molenda[2], I. Moskal[1], Sz. Niedźwiecki[1], M. Pałka[1], M. Pawlik[1], L. Raczyński[3], Z. Rudy[1], P. Salabura[1], N.G. Sharma[1], M. Silarski[1], A. Słomski[1], J. Smyrski[1], A. Strzelecki[1], K. Szymanski[1], W. Wiślicki[3], P. Witkowski[1], M. Zieliński[1], N. Zoń[1]

[1] Institute of Physics, Jagiellonian University, 30-059 Kraków, Poland

[2] Faculty of Chemistry, Jagiellonian University, 30-060 Kraków, Poland

[3] Świerk Computing Centre, National Centre for Nuclear Research, 05-400 Otwock-Świerk, Poland

## Abstract

Photomultipliers are commonly used in commercial PET scanner as devices which convert light produced in scintillator by gamma quanta from positron-electron annihilation into electrical signal. For proper analysis of obtained electrical signal, a photomultiplier gain curve must be known, since gain can be significantly different even between photomultipliers of the same model. In this article we describe single photoelectron method used for photomultipliers calibration applied for J-PET scanner, a novel PET detector being developed at the Jagiellonian University. Description of calibration method, an example of calibration curve and gain of few R4998 Hamamatsu photomultipliers are presented.

## 1. Introduction

Positron emission tomography (PET) allows to imagine spatial distribution of injected substance and temporal changes of biological and chemical processes inside the body. It is used in medical diagnostics (e.g. oncology, cardiology) and to monitor changes in brain in case of people with mental diseases. Tomographs (which are currently commercially used) consist of hundreds of scintillating crystals which detect gamma radiation coming from decay of radioactive pharmaceuticals injected into patient's bloodstream. Scintillating crystals are usually grouped in blocks with dimensions up to 5cm x 5cm. At the rear of each block typically four photomultipliers convert light pulses into electrical signals. Reconstruction of images is mainly based on signals coming from the photoelectric effect in scintillating crystals [1].

Calibration check of PET device with radioactive source is done every day before patients examination to control if all photomultipliers are set to the same gain value. If needed, a change of applied voltage must be done according to individual calibration curve.

## 2. Definition of photomultiplier tube gain

A photon which hit a photomultiplier's photocathode, can release a photoelectron. The emitted photoelectron is accelerated in an electric field inside photomultiplier. Thus after striking the first dynode secondary electrons emission occur. The secondary emissions are repeated over all dynodes resulting in a high current amplification (Fig. 1). Secondary electron emission ratio δ between two dynodes is given by $\delta = A \cdot E^\alpha$, where A is a constant, E is an interstage voltage, and α is a coefficient depending on the dynodes and the geometric structure. In a photomultiplier with n dynodes gain μ becomes

$$\mu = \delta^n = (A \cdot E^\alpha)^n = K \cdot V^{\alpha n}, \quad (1)$$

where K is a constant and V is a voltage applied between cathode and anode [2].

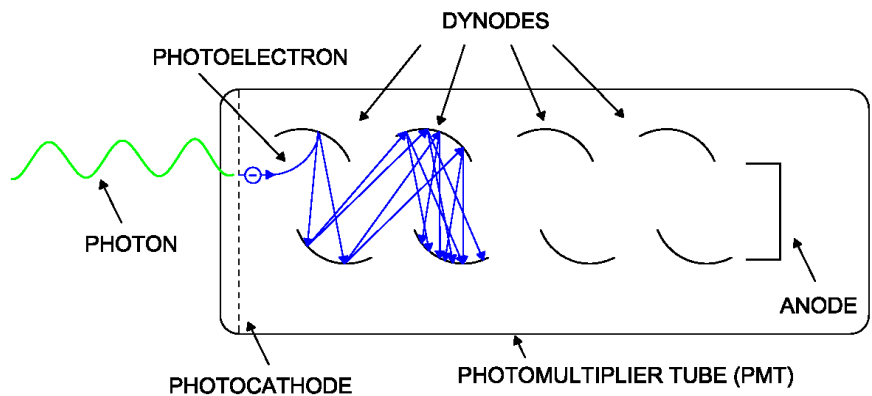

**Fig. 1.** A photomultiplier tube scheme with secondary electron emissions.

For a given amount of light reaching the photocathode at the applied voltage the charge of produced electric signal is proportional to the gain of a given photomultiplier. However gain of photomultipliers can be significantly different even for the same model of photomultiplier as it is shown in Fig. 2 (Hamamatsu R4998 photomultipliers [2] with 2250 V applied, optimal for the model).

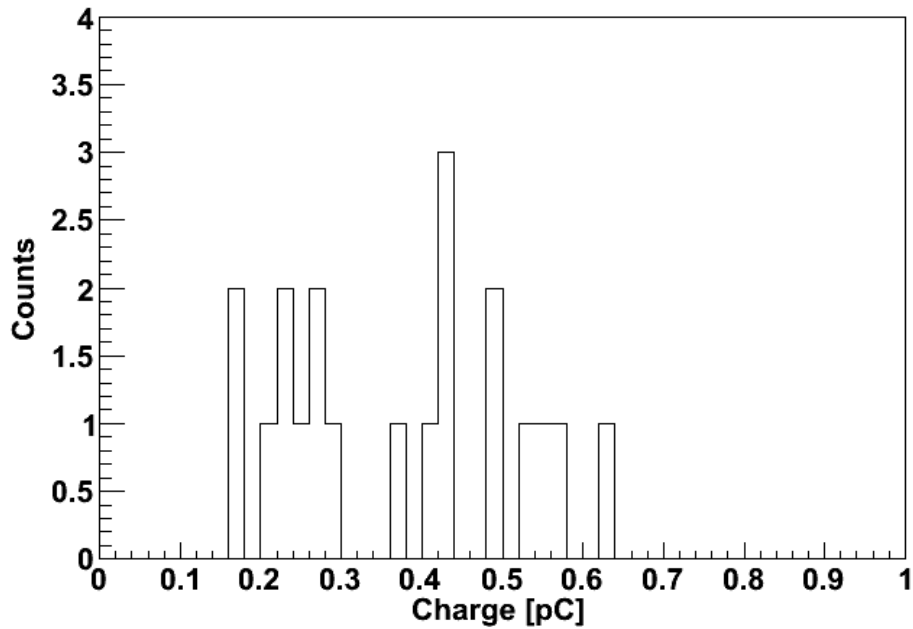

**Fig. 2.** Single photoelectron charge for 20 Hamamatsu R4998 photomultipliers with applied high voltage of 2250 V.

Thus for proper functionality of PET devices photomultipliers gain calibration is important.

## 3. Calibration of single photomultiplier tube

One of the methods of determination of the calibration curve of photomultiplier tube (PMT) is to measure a charge of the PMT signal when a particular amount of light is directed on a photomultiplier's photocathode, therefore pulse lasers are commonly used as the light sources. The other possibility is to monitor the light intensity of the source with proper device e.g. avalanche photodiode [3]. A different approach is to measure charge from a PMT when only single photon react with the photocathode. If light intensity is low, then it may happen that zero or only one of photons interact with the photocathode releasing photoelectron. To obtain the low intensity of light the additional attenuation can be introduced [4]. A number of photoelectrons can be also estimated from observed signals produced by a radioactive source of known radiation energy when relation between number of released photoelectrons and energy of the observed signal is known [5].

In this article we present a calibration method based on the single photoelectron approach used for J-PET at Jagiellonian University [6], with modifications mentioned in the next sections.

## 4. Single photoelectron method

Gamma quanta going through the scintillating material can excite medium atoms (the scintillator used during the calibration was RP-422 produced by Rexon with light output of about 8400 photons per 1 MeV of deposited energy [7], [8]). During the atoms de-excitation photons are isotropically emitted. Some of them may fly away from the scintillator but others can be internally reflected and reach its end, where a photomultiplier is attached. Two methods can be applied in order to have reaction of only one photon with the PMT photocathode from a hit of a single gamma quantum in the scintillator: (i) the light output of the scintillator must be attenuated or (ii) an aperture between the photomultiplier and the scintillator is inserted to restrict area of the photocathode which can be hit by photons. The second method is in use for the J-PET project.

If we assume that the probability to produce a photoelectron in the photocathode is the same for each photon reaching the edge of the scintillator then the probability distribution of the number of photoelectrons is given by the Bernoulli distribution:

$$P_k^{N_{ph}} = \binom{N_{ph}}{k} p^k (1-p)^{N_{ph}-k}, \quad (2)$$

where $N_{ph}$ is a number of photons reaching the end of the scintillator, k is a number of photoelectrons released from the photocathode and p is a probability that a photon reaching the edge of the scintillator will cause emission of photoelectron. This probability can be defined as follows

$$p = \varepsilon \cdot \frac{A_{aperture}}{A_{scintillator}}, \quad (3)$$

where A is area of an aperture (0.28 $mm^2$) and a scintillator cross-section (196 $mm^2$), $\varepsilon$ is a cathode quantum efficiency (which is about 20% for used Hamamatsu R4998 PMT model [2]). For the above mentioned values $p = 2.88 \cdot 10^{-4}$. Based on Eq. 2 probability for 0, 1 and 2 photoelectrons can be written as:

$$P_0^{N_{ph}} = (1-p)^{N_{ph}}, \quad (4)$$

$$P_1^{N_{ph}} = N_{ph} \cdot p \cdot (1-p)^{N_{ph}-1}, \quad (5)$$

$$P_2^{N_{ph}} = \frac{1}{2} N_{ph} \cdot (N_{ph}-1) \cdot p^2 (1-p)^{N_{ph}-2}. \quad (6)$$

Based on Eq. 4 – 6 the ratio between number of events with zero, one and two photoelectrons can be calculated as:

$$\frac{N_0}{N_1} = \frac{1-p}{N_{ph} \cdot p}, \quad (7)$$

$$\frac{N_1}{N_2} = \frac{2(1-p)}{(N_{ph}-1) \cdot p}. \quad (8)$$

Dependence of ratios from Eq. 7 and 8 on probability p is shown in Fig. 3.

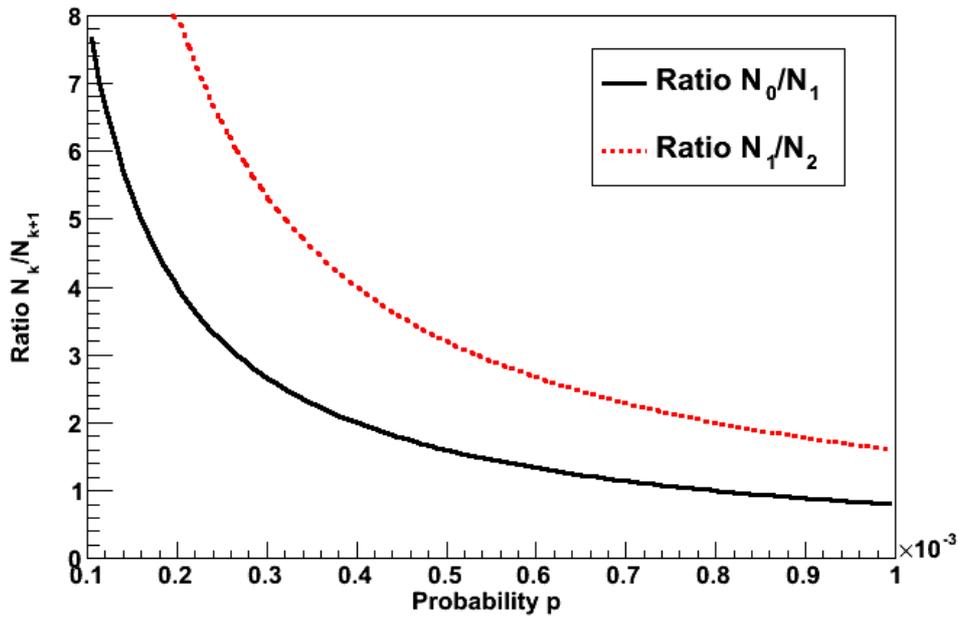

**Fig. 3.** Calculated ratios of consecutive number of observed photoelectrons as a function of probability of releasing photoelectron from a photocathode.

The number of photons used to obtain ratios in Fig. 3 was derived from the average signal charge spectrum from the reference photomultiplier (mentioned in the next section) and is $N_{ph} = 1250$. From Fig. 3 we can infer that ratio of number of events with one and two photoelectrons should be about two times larger than for ratio of events with zero and one photoelectron.

## 5. Measurement setup and data collection

The measurement setup consists of two Hamamatsu photomultipliers R4998 [2], a Rexon scintillator RP422 [7] with dimensions 14 x 14 x 100 mm$^3$, a $^{22}$Na radioactive source and LeCroy WaveRunner 64Xi oscilloscope [9]. Between the scintillator and the tested photomultiplier an aperture with a hole diameter of 0.6 mm was inserted (Fig. 4).

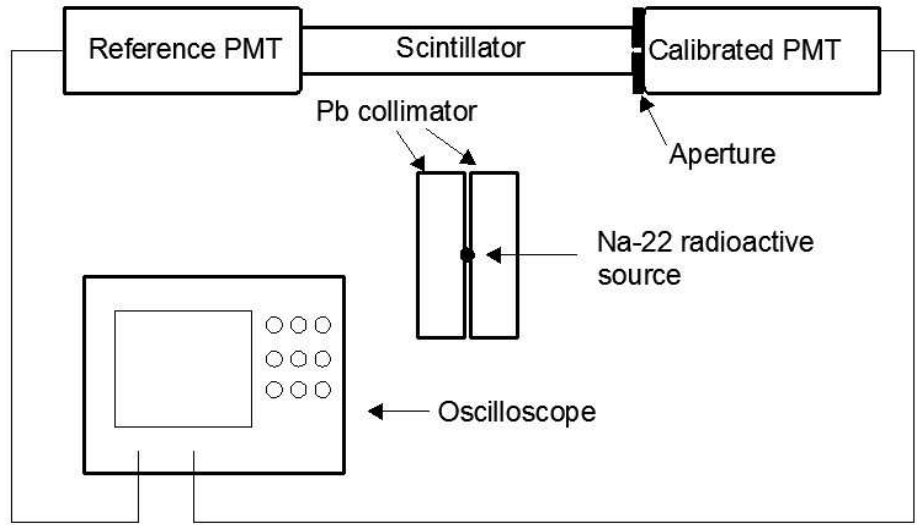

**Fig. 4.** A scheme of measurement setup.

The second photomultiplier was used as a reference detector, while the oscilloscope was used to collect the complete shape of each signal. Signals from the calibrated photomultiplier were collected only when the reference PMT gave a signal in order to avoid collecting a photomultiplier thermal noise. A distribution of signals charge from the calibrated photomultiplier is shown in Fig. 5.

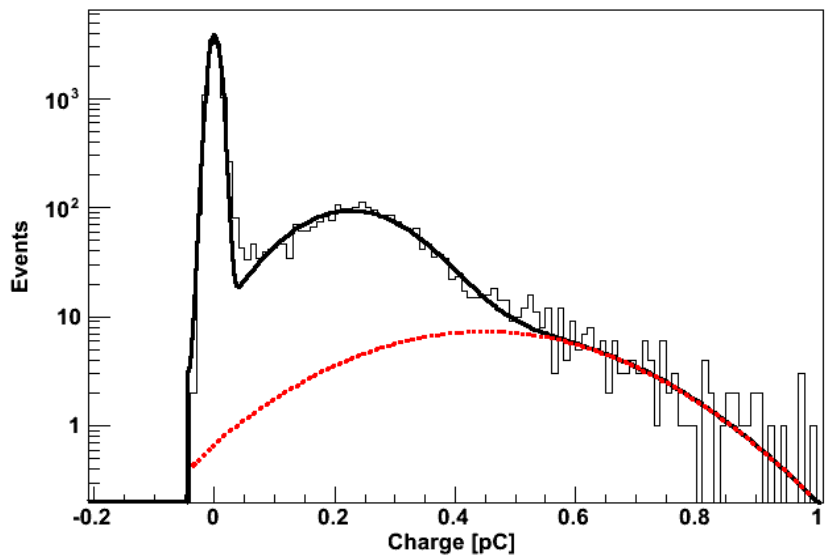

**Fig. 5**. A histogram of charge distribution collected during the measurement. Fitted curve is a sum of three Gaussian functions. From the left side: around 0 pC a case when no photoelectron was emitted, single (around 0.25 pC) and two (around 0.5 pC) photoelectrons maximum. The red dashed, line is separately drawn for two photoelectrons maximum.

In the distribution in Fig. 5 one can distinguish three maxima. From the left side: a maximum from events when no photoelectron was released from a photocathode, a maximum from a one photoelectron and a maximum from two photoelectrons. The two photoelectrons maximum shows up as a bump on the right side of the one photoelectron maximum. Fitted curve F is a sum of three Gaussian functions given as:

$$F = N_0 \cdot \exp\left(\frac{-(x - X_0^0)^2}{2\sigma_0^2}\right) + N_1 \cdot \exp\left(\frac{-(x - X_1^0)^2}{2\sigma_1^2}\right) + N_2 \cdot \exp\left(\frac{-(x - X_2^0)^2}{2\sigma_2^2}\right), \quad (9)$$

where $N_x$, $X_x^0$, $\sigma_x$ (x=0,1,2) correspond to normalizations, centre of maxima and standard deviations, respectively.

Due to the fact that the gathered charge should scale linearly with the number of observed photoelectrons, the position of the centre of two photoelectrons maximum is, as expected, two times larger than for one photoelectron. Therefore, in order to decrease number of fitted parameters Eq. 9 can be rewritten:

$$F = N_0 \cdot \exp\left(\frac{-(x - X_0^0)^2}{2\sigma_0^2}\right) + N_1 \cdot \exp\left(\frac{-(x - X_1^0)^2}{2\sigma_1^2}\right) + N_2 \cdot \exp\left(\frac{-(x - 2X_1^0)^2}{2\sigma_2^2}\right). \quad (10)$$

Values of fitted parameters allows for determination of ratios from Eq. 7 and 8. To compare the experimental results with the predictions one can use Eq. 8 to perform an estimation of $N_{ph}$. For ratio obtained from Fig. 5, ($N_1/N_2 = 6.89$) and the known value of $p = 2.88 \cdot 10^{-4}$, one obtains $N_{ph}$ = 1139 photons, which is in good agreement with the value derived from the reference detector $N_{ph}$ =1250.

## 6. Calibration curve

The information about charge induced in a photomultiplier from a single photoelectron allows for determination of a PMT gain curve. The gain is changing with the voltage applied on the PMT (Eq. 1). Therefore measurement of the charge given by the PMT for the single photoelectron events as a function of voltage applied to it allows for determination of the gain curve. An example of such curve is shown in Fig. 6. The curve (black solid line) of a form given in Eq. 1 is fitted to the experimental data points.

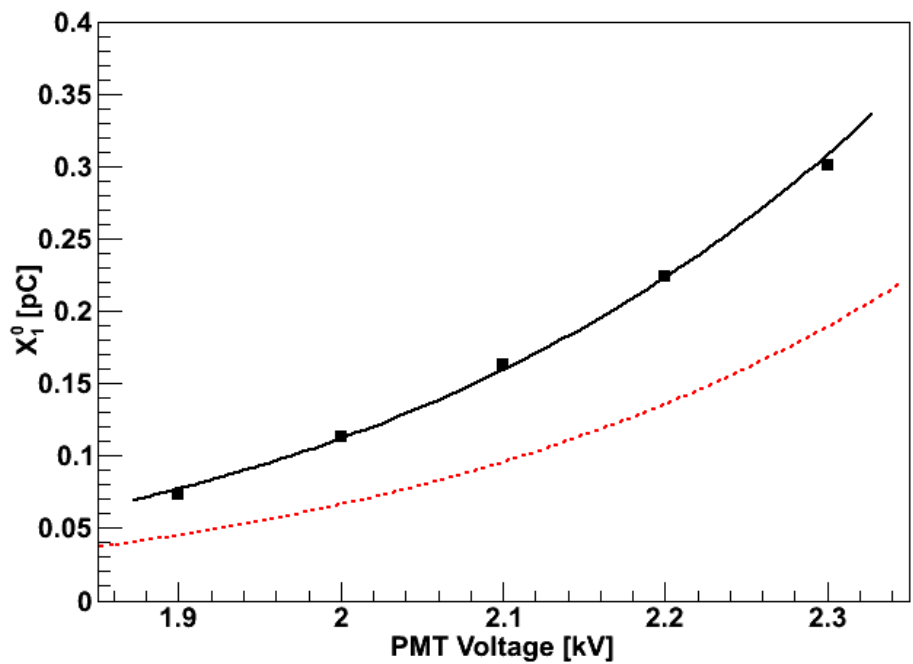

**Fig. 6.** A power function (black solid line) of form $A \cdot x^B$ fitted to experimental data points. Derived parameters: $A = (7.31 \pm 0.43) \cdot 10^{-4}\ [pC / kV^B]$ and $B = 7.252 \pm 0.098$. The experimental errors are smaller than presented data points. For comparison gain curve (red dashed line) for the other photomultiplier is shown.

In Fig. 6 additional gain curve (for other photomultiplier) is shown as red dashed line. To set the same value of gain (charge induced for single photoelectron about 0.16 pC = $10^4$ e) on both photomultipliers voltage of about 2100 V and 2250 V has to be applied on first and second one, respectively.

## 7. Summary

The single photoelectron method was described which enables to determine the gain curve for photomultiplier needed for calculation of a number of photoelectrons in observed signal. The number of photoelectrons is more convenient to use in comparisons of signals instead of signal charge, since it does not depend on voltage applied to the photomultiplier.

Gain of photomultipliers can differ significantly even for the same model therefore gain calibration is necessary for proper operation of PET devices.

## Acknowledgements

We acknowledge technical and administrative support by M. Adamczyk, T. Gucwa-Ryś, A. Heczko, M. Kajetanowicz, G. Konopka-Cupiał, J. Majewski, W. Migdał, A. Misiak, and the financial support by the Polish National Center for Development and Research through grant INNOTECH-K1/IN1/64/159174/NCBR/12, the Foundation for Polish Science through MPD programme, the EU and MSHE Grant No. POIG.02.03.00-161 00-013/09 and Małopolskie Centre of Entrepreneurship through Doctus programme.